\documentclass[aps,prb,twocolumn]{revtex4}

\usepackage{graphicx}
\usepackage{amssymb,amsmath}

\usepackage{color}
\usepackage{epsfig}
\usepackage{psfrag}
\usepackage{epstopdf}

\makeatletter
\def\@dotsep{4.5}
\makeatother

\begin{document}
\newcommand{\commute}[2]{\left[#1,#2\right]}
\newcommand{\bra}[1]{\left\langle #1\right|}
\newcommand{\ket}[1]{\left|#1\right\rangle }
\newcommand{\anticommute}[2]{\left\{  #1,#2\right\}  }

\title{Quantum vs. classical hyperfine-induced dynamics in a quantum dot}

\author{W.~A. Coish}
\affiliation{Department of Physics and Astronomy, University of Basel, Klingelbergstrasse 82, 4056 Basel, Switzerland}
\author{E.~A. Yuzbashyan}
\affiliation{Center for Materials Theory, Department of Physics and Astronomy, Rutgers University, Piscataway, New Jersey 08854, USA}
\author{B.~L. Altshuler}
\affiliation{Physis Department, Columbia University, 538 West 120th Street, New York, NY 10027, USA}
\affiliation{NEC-Laboratories America, 4 Independence Way, Princeton, New Jersey 085540, USA}
\author{Daniel Loss}
\affiliation{Department of Physics and Astronomy, University of Basel, Klingelbergstrasse 82, 4056 Basel, Switzerland}

\date{\today}

 
\begin{abstract}
In this article we analyze spin dynamics for electrons confined to semiconductor quantum dots due to the contact hyperfine interaction. We compare mean-field (classical) evolution of an electron spin in the presence of a nuclear field with the exact quantum evolution for the special case of uniform hyperfine coupling constants.  We find that (in this special case) the zero-magnetic-field dynamics due to the mean-field approximation and quantum evolution are similar.  However, in a finite magnetic field, the quantum and classical solutions agree only up to a certain time scale $t<\tau_c$, after which they differ markedly.     
\end{abstract}
\maketitle                   

\section{Introduction}
Prospects for future quantum information processing with quantum-dot-confined electron spins\cite{loss:1998a} have encouraged a series of recent experimental efforts. These efforts have resulted in several very significant achievements, including single-electron confinement in vertical\cite{tarucha:1996a} and lateral single\cite{ciorga:2000a} and double\cite{elzerman:2003a,petta:2004a} gated quantum dots, the demonstration of spin-dependent transport in double dots,\cite{ono:2002a, ono:2004a, koppens:2005a} and exciting effects arising from the contact hyperfine interaction with nuclear spins in the host material, including coherent undriven oscillations in spin-dependent transport,\cite{ono:2004a} lifting of the spin-blockade,\cite{koppens:2005a} enhancement of the nuclear spin decay rate near sequential-tunneling peaks,\cite{lyanda-geller:2002a,huettel:2004a} and notably, decay of coherent oscillations between singlet and triplet states as well as the demonstration of two-qubit gates in double quantum dots.\cite{petta:2005a,laird:2005a}  Very recently, the hyperfine interaction has also been identified as the source of decay for driven single-spin Rabi oscillations in quantum dots.\cite{koppens:2006a, koppens:2007a} 

In spite of rapid progress, there are still many obstacles to quantum computing with quantum dots.  In particular, the inevitable loss of qubit coherence due to fluctuations in the environment is acceptable in a quantum computer only if the error rates due to this  loss are kept below $10^{-3}-10^{-4}$ errors per operation.\cite{steane:2003a}  This requirement is particularly difficult to achieve since it means that interactions must be strong while switching so that operations can be performed rapidly, but still very weak in the idle state, to preserve coherence.

For an electron spin confined to a quantum dot, decoherence can proceed through fluctuations in the electromagnetic environment and spin-orbit interaction\cite{khaetskii:2000a,khaetskii:2001a, golovach:2004a,bulaev:2005a,borhani:2005a} or through the hyperfine interaction with nuclei in the surrounding host material, which has been shown extensively in theory\cite{burkard:1999a,erlingsson:2001a,khaetskii:2002a,khaetskii:2003a,erlingsson:2002a,merkulov:2002a,schliemann:2002a,schliemann:2003a,desousa:2003a,desousa:2003b,coish:2004a,abalmassov:2004a,erlingsson:2004a,desousa:2005a,shenvi:2005a,shenvi:2005b,taylor:2005a,taylor:2006a,yao:2005a,witzel:2005a,coish:2005a,klauser:2006a,yao:2006a}  and experiment.\cite{bracker:2005a, dutt:2005a, johnson:2005a,johnson:2005b,petta:2005a,petta:2005b,laird:2005a}  Due to the primarily $p$-type nature of the valence band in GaAs, hole spins (unlike electron spins) do not couple to the nuclear spin environment via the contact hyperfine interaction, although they can still undergo decay due to spin-orbit coupling. The decay may still occur on an even longer time scale than for electrons,\cite{bulaev:2005b} which suggests the dot-confined hole spin may be another good candidate for quantum computing.  Alternatively, quantum dots fabricated in isotopically purified $^{28}\mathrm{Si}$\cite{friesen:2003a} or $^{12}\mathrm{C}$ nanotubes\cite{mason:2004a,sapmaz:2006a,graeber:2006a} would be free of nuclei with spin, and therefore free of hyperfine-induced decoherence. 

While the field of quantum-dot spin decoherence has been very active in the last few years, there still remain significant misconceptions regarding the nature of the most relevant (hyperfine) coupling, particularly, the range of validity of semiclassical spin models and traditional decoherence methods involving ensemble averaging have been called into question for a single isolated quantum dot with a potentially controllable environment.  We address these issues in section \ref{sec:hyperfine}.

\section{Hyperfine interaction: quantum and classical dynamics}
\label{sec:hyperfine}
Exponential decay of the longitudinal and transverse components of spin is typically measured by the decay time scales $T_1$ and $T_2$, respectively.\cite{slichter:1980a}  The longitudinal spin relaxation rate $1/T_1$ due to spin-orbit interaction and phonon emission is significantly reduced in quantum dots relative to the bulk in the presence of a weak Zeeman splitting $B=|\mathbf{B}|$ and large orbital level spacing $\hbar\omega_0$ ($1/T_1\propto B^5/(\hbar\omega_0)^4$).\cite{khaetskii:2001a, golovach:2004a}  This decay time has been shown to be on the order of $T_1\sim1\,\mathrm{ms}$ in gated GaAs quantum dots at $B\approx8\,\mathrm{T}$,\cite{elzerman:2004a} and to reach a value as large as $T_1=170\,\mathrm{ms}$ at low magnetic fields ($B=1.75\,\mathrm{T}$).\cite{amasha:2006a} Furthermore, since dephasing is absent at leading order for fluctuations that couple through the spin-orbit interaction, the $T_2$ time due to this mechanism is limited by the $T_1$ time ($T_2=2T_1$)\cite{golovach:2004a} (we note that corrections at higher order in the spin-orbit interaction can lead to pure dephasing, although these corrections are only relevant at very low magnetic fields \cite{san-jose:2006a,coish:2006b}). Unlike the spin-orbit interaction, the hyperfine interaction \emph{can} lead to pure dephasing of electron spin states at leading order, resulting in a relatively very short decoherence time $\tau_\mathrm{c}\approx1-10\,\mathrm{ns}$ due to non-exponential (Gaussian) decay.\cite{khaetskii:2002a,merkulov:2002a}  To perform quantum-dot computations, this and any additional decay must be fully understood and reduced, if possible.

The Hamiltonian for an electron spin $\mathbf{S}$ in the lowest orbital level of a quantum dot containing nuclear spins is 
\begin{equation}
H_\mathrm{hf}=\mathbf{S}\cdot\left(\mathbf{B}+\mathbf{h}\right);\,\,\,\,\,\mathbf{h}=\sum_i A_i\mathbf{I}_i, \label{eq:HFHamiltonian}
\end{equation}
where $A_i=Av_0\left|\psi_0(\mathbf{r}_i)\right|^2$ is the contact hyperfine coupling constant to the nuclear spin at site $i$, $v_0$ is the volume of a crystal unit cell containing one nuclear spin, and $A\approx 90\,\mu eV$ is the weighted average hyperfine coupling constant in GaAs, averaged over the coupling constants for the three naturally occurring radioisotopes $^{69}\mathrm{Ga},^{71}\mathrm{Ga}$, and $^{75}\mathrm{As}$ (weighted by their natural abundances),\cite{paget:1977a} all with total nuclear spin $I=3/2$.  The nuclear field in $H_\mathrm{hf}$ is given by the quantum ``Overhauser operator" $\mathbf{h}$. Although an exact Bethe Ansatz solution exists for $H_\mathrm{hf}$,\cite{gaudin:1976a} using this solution to perform calculations for the full coupled quantum system of $N\approx10^4-10^6$ nuclei and one electron in a quantum dot can be prohibitively difficult.\cite{schliemann:2003a}  Since the Overhauser operator $\mathbf{h}$ is a sum of a large number $N$ of spin-$I$ operators, one expects that under certain conditions its quantum fluctuations can be neglected and the operator $\mathbf{h}$ can be replaced with a classical Overhauser field $\mathbf{h}\to\mathbf{B}_N$.\cite{schulten:1978a,khaetskii:2002a,merkulov:2002a,semenov:2003a,erlingsson:2004a,erlingsson:2005b,bracker:2005a, braun:2005a, dutt:2005a,yuzbashyan:2005a, taylor:2005a, taylor:2005b, taylor:2005c, taylor:2006a, petta:2005a, koppens:2005a,jouravlev:2006a}  However, this approximation can accurately describe the electron-spin dynamics only at times $t<\tau_c$, where $\tau_c= N^\eta/A$ and $\eta>0$\cite{yuzbashyan:2005a},\footnote{This expression, of course, assumes an appropriate scaling of coupling constants $A_i\propto 1/N$, so that the energy of the electron spin scales as $N^0$ in the thermodynamic limit.} after which effects of quantum fluctuations of the Overhauser operator set in.  The nuclei in GaAs are indeed quantum objects, which could be verified, in principle, by demonstrating that they can be entangled, as is done in spin-state squeezing experiments that have been performed on atomic ensembles.\cite{geremia:2004a}  The replacement $\mathbf{h}\to\mathbf{B}_N$ is therefore not exact and there are several cases in which the electron-spin dynamics at times $t>\tau_c$ differ markedly for quantum and classical nuclear fields. In particular, without performing an ensemble average over initial Overhauser fields, the classical-field picture predicts no decay of the electron spin.  This is in direct contradiction to analytical \cite{coish:2004a,zurek:2005a,coish:2005a,klauser:2006a} and exact numerical \cite{schliemann:2002a, shenvi:2005a} studies that show the quantum nature of the nuclei can lead to complete decay of the transverse electron spin, even in the presence of a static environment (fixed initial conditions).  Additionally, quantum ``flip-flop'' processes can lead to dynamics and decay of the electron spin in the quantum problem, even for initial conditions (e.g., a fully-polarized nuclear spin system) that correspond to a fixed-point of the classical equations of motion.\cite{khaetskii:2002a,coish:2004a,shenvi:2005b}  In fact, it can be shown that any decay of the electron spin for pure-state inital conditions will result in quantum entanglement between the electron and nuclear spin systems.\cite{schliemann:2002a,schliemann:2003a}  This entanglement has recently been highlighted as a source of spin-echo envelope decay in the presence of the hyperfine interaction.\cite{yao:2006a}  Finally, even the ensemble-averaged standard classical (mean-field) electron-spin dynamics show large quantitative differences relative to the exact quantum dynamics at times $t>\tau_c$ and in a very weak magnetic field, although an alternative mean-field theory involving the P-representation for the density matrix shows promise.\cite{alhassanieh:2005a}           

While the classical and quantum dynamics diverge in many cases, the classical-field replacement $\mathbf{h}\to\mathbf{B}_N$ \emph{will} be valid up to some time scale, providing a range of validity for the classical dynamics.  In this article, we aim to shed light on this range of validity of the classical solution.  As a test of the classical-dynamics picture, we can compare quantum and classical dynamics of an electron spin in the simple case of uniform coupling constants $A_i=\gamma$. When the coupling constants are uniform, an exact solution to the quantum dynamics (see Refs. [\onlinecite{khaetskii:2003a,eto:2004a}] for the $\left|\mathbf{B}\right|=0$ case) can be evaluated and used to compare with an integration of the equivalent classical equations of motion. For uniform coupling constants, the nuclear Overhauser operator from Eq. (\ref{eq:HFHamiltonian}) becomes $\mathbf{h}=\gamma\mathbf{K}$, where $A_i = \gamma = A/N$ and 
$\mathbf{K}=\sum_i\mathbf{I}_i$ is the collective total spin operator for $N\gg1$ nuclear spins.   

The initial state of the system is taken to be an arbitrary product
state of the electron and nuclear system:
\begin{eqnarray}
\ket{\psi(0)} & = & \ket{\psi_{S}(0)}\otimes\ket{\psi_{K}(0)}, \\
              & = & \sum_{m=-K}^{K}\left(\alpha_{m}^{\uparrow}\ket{\uparrow;K,m}+\alpha_{m}^{\downarrow}\ket{\downarrow;K,m}\right),\label{eq:ICs} 
\end{eqnarray}
where $\ket{\sigma;K,m}$ is a simultaneous eigenstate of $S_z$, $K_z$ (we take the direction of the external field $\mathbf{B}$ to define the $z$-axis), and $\mathbf{K}\cdot\mathbf{K}$ (with eigenvalues $\pm1/2$ for $\sigma=\uparrow,\downarrow$, $m$, and $K(K+1)$, respectively).  For comparison with the classical spin dynamics, we choose the collective nuclear spin to be initially described by a spin coherent state, given by $\ket{\psi_{K}(0)}=e^{-iK_{y}\theta_{K}}\ket{K,K}=\sum_{m}d_{mK}^{(K)}(\theta_{K})\ket{K,m}$,
where $d_{mK}^{(K)}(\theta_{K})$ is the Wigner rotation matrix\cite{sakurai_wignersymbol:1985a} and the electron spin is in an arbitrary initial state $\ket{\psi_{S}(0)}=\cos(\theta_{S}/2)\ket{\uparrow}+e^{i\phi_{S}}\sin(\theta_{S}/2)\ket{\downarrow}$.  The initial conditions are then completely determined by the three angles $\theta_{S},\phi_{S}$, and $\theta_{K}$.  These initial conditions allow for an arbitrary relative orientation of the spin and magnetic-field vectors, since the azimuthal angle for $\mathbf{K}$ ($\phi_K$) can be set to zero with an appropriate shift in $\phi_S$: $\phi_K^\prime=0,\,\,\phi_S^\prime=\phi_S-\phi_K$. At any later time $t$, the wave function is given by
\begin{equation}
\ket{\psi(t)}=\sum_{m=-K}^{K}\left(\alpha_{m}^{\uparrow}(t)\ket{\uparrow;K,m}+\alpha_{m}^{\downarrow}(t)\ket{\downarrow;K,m}\right).\label{eq:Psioft}
\end{equation}

\begin{figure*}[htb]
\includegraphics[width=\textwidth]{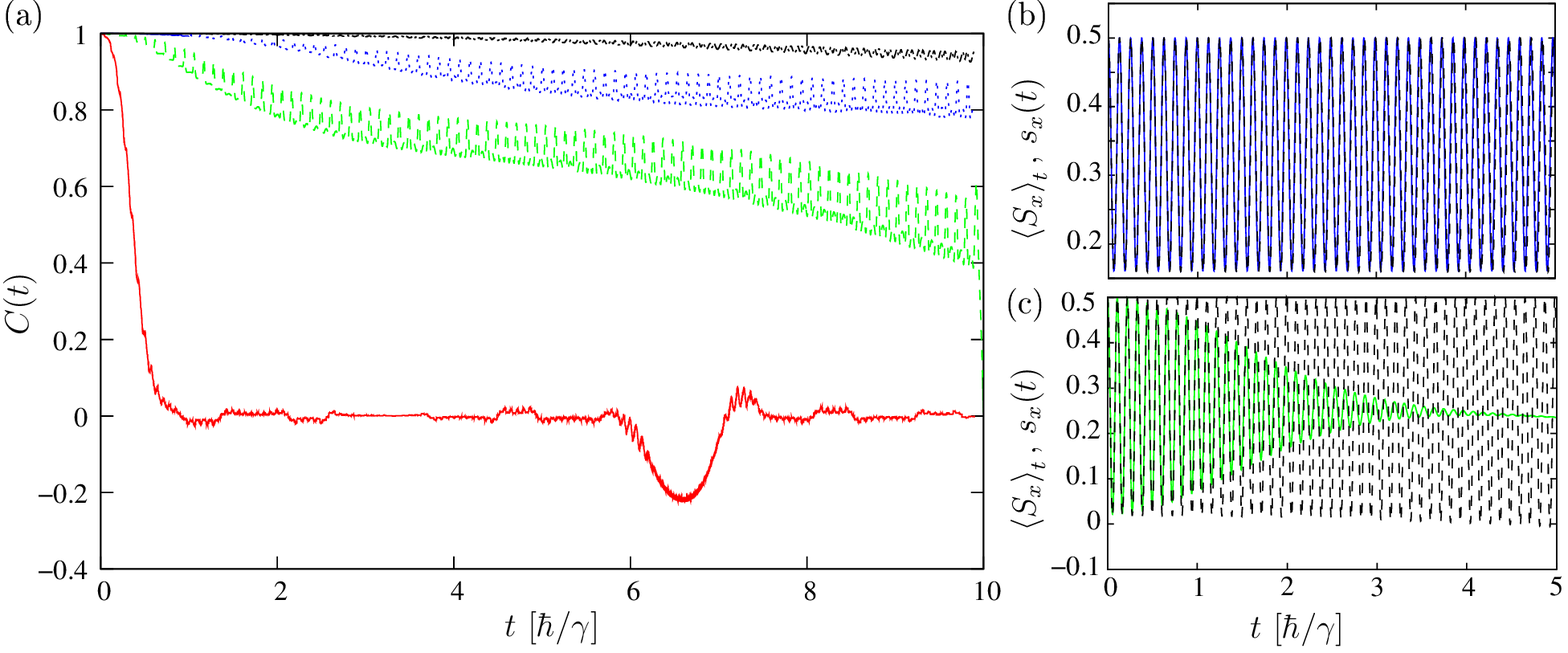}
\caption{(Color online) (a) Correlation between the mean-field and exact quantum solution $C(t)$ (defined in Eq. (\ref{eq:CDefinition}); $C(t)=1$ indicates perfect agreement between the mean-field and quantum solutions) for evolution of an electron spin in the presence of a total bath spin $K=50$ and magnetic field $B=0$ (black dash-dotted line, showing the weakest decay), $B=5\gamma$ (blue dotted line), $B=10\gamma$ (green dashed line) and large magnetic field $B=500\gamma$ (red solid line, showing rapid decay).  The inital conditions were $\theta_S = \pi/2$, $\phi_S = 0$, $\theta_K=0.3 \pi$ (see the discussion following Eq. (\ref{eq:ICs})). We also show the exact quantum evolution $\left<S_x\right>_t$ (solid line) and mean-field approximation $s_x(t)$ (dashed line) for (b) $B=0$ and (c) $B=10\gamma$.}
\label{fig:MFandQuantumDynamics}
\end{figure*}
 From the time-dependent Schr\"odinger equation $i\partial_{t}\ket{\psi(t)}=H_{\mathrm{hf}}\ket{\psi(t)}$ (setting $\hbar=1$), we find the set of coupled differential equations determining the coefficients $\left\{ \alpha_{m}^{\uparrow}(t),\alpha_{m}^{\downarrow}(t)\right\} $. For $m=-K...K-1$,
\begin{eqnarray}
\dot{\alpha}_{m}^{\uparrow} & = & -\frac{i}{2}(B+\gamma m)\alpha_{m}^{\uparrow}-i\frac{\gamma}{2}C_{Km+1}^{-}\alpha_{m+1}^{\downarrow},\label{eq:AlphaDiffEq1}\\
\dot{\alpha}_{m+1}^{\downarrow} & = & \frac{i}{2}(B+\gamma(m+1))\alpha_{m+1}^{\downarrow}-i\frac{\gamma}{2}C_{Km}^{+}\alpha_{m}^{\uparrow},\label{eq:AlphaDiffEq2}
\end{eqnarray}
where $C_{Km}^{\pm}=\bra{Km\pm1}K_{\pm}\ket{Km}=\sqrt{K(K+1)-m(m\pm1)}$.
These equations are supplemented by two equations for the stationary
states $\ket{\uparrow;K,K}$ and $\ket{\downarrow;K,-K}$ with dynamics:
\begin{eqnarray}
\alpha_{K}^{\uparrow}(t) & = & \exp\left\{-\frac{i}{2}(B+\gamma K)t\right\} \alpha_{K}^{\uparrow}(0),\label{eq:alphaK}\\
\alpha_{-K}^{\downarrow}(t) & = & \exp\left\{ \frac{i}{2}(B-\gamma K)t\right\} \alpha_{-K}^{\downarrow}(0).\label{eq:alphamK}
\end{eqnarray}
The solutions to Eqs. (\ref{eq:AlphaDiffEq1}), (\ref{eq:AlphaDiffEq2}) and the expressions in Eqs. (\ref{eq:alphaK}), (\ref{eq:alphamK}) for the coefficients $\{\alpha_{m}^{\uparrow}(t),\alpha_{m}^{\downarrow}(t):m=-K\ldots K\}$
constitute a complete exact solution for the dynamics of the wave
function $\ket{\psi(t)}$ at any later time $t>0$. We solve Eqs. (\ref{eq:AlphaDiffEq1}) and (\ref{eq:AlphaDiffEq2}) by Laplace transformation to obtain
\begin{widetext}
\begin{eqnarray}
\alpha_{m}^{\uparrow}(t) & = & e^{i\frac{\gamma}{4}t}\left\{ \alpha_{m}^{\uparrow}(0)\cos(\omega_{Km}t)-i\left(\alpha_{m}^{\uparrow}(0)\left[B+\gamma(m+\frac{1}{2})\right]+\alpha_{m+1}^{\downarrow}(0)\gamma C_{Km+1}^{-}\right)\frac{\sin(\omega_{Km}t)}{2\omega_{Km}}\right\} ,\label{eq:alpham}\\
\alpha_{m+1}^{\downarrow}(t) & = & e^{i\frac{\gamma}{4}t}\left\{ \alpha_{m+1}^{\downarrow}(0)\cos(\omega_{Km}t)+i\left(\alpha_{m+1}^{\downarrow}(0)\left[B+\gamma(m+\frac{1}{2})\right]-\alpha_{m}^{\uparrow}(0)\gamma C_{Km}^{+}\right)\frac{\sin(\omega_{Km}t)}{2\omega_{Km}}\right\} ,\label{eq:alphamp1}\\
\omega_{Km} & = & \frac{1}{2}\left[(B+\gamma m)(B+\gamma(m+1))+\gamma^{2}\left(C_{Km+1}^{-}C_{Km}^{+}+\frac{1}{4}\right)\right]^{1/2}.
\end{eqnarray}
\end{widetext}
With the coefficients $\{\alpha_{m}^{\uparrow}(t),\alpha_{m}^{\downarrow}(t):m=-K\ldots K\}$ in hand, we can evaluate the expectation values of all spin components exactly: $\left\langle \mathbf{S}\right\rangle _{t}=\bra{\psi(t)}\mathbf{S}\ket{\psi(t)},\left\langle \mathbf{K}\right\rangle _{t}=\bra{\psi(t)}\mathbf{K}\ket{\psi(t)}$.

To evaluate the classical spin dynamics, we perform a mean-field decomposition of the Hamiltonian given in Eq.
(\ref{eq:HFHamiltonian}) by rewriting the spin operators as $\mathbf{S}=\left\langle \mathbf{S}\right\rangle _{t}+\delta\mathbf{S}$
and $\mathbf{K}=\left\langle \mathbf{K}\right\rangle _{t}+\delta\mathbf{K}$.
We then neglect the term that is bilinear in the spin fluctuations 
($\propto \delta\mathbf{S}\cdot\delta\mathbf{K}$) and approximate the spin expectation values by their self-consistent mean-field dynamics $\left<\mathbf{S}\right>_t\approx\mathbf{s}(t)$, $\left<\mathbf{K}\right>_t\approx\mathbf{k}(t)$, where $\mathbf{s}$ and $\mathbf{k}$ are classical time-dependent vectors of fixed length.\cite{yuzbashyan:2005a}
Up to a c-number shift, this gives the (time-dependent) mean-field
Hamiltonian\begin{equation}
H_{\mathrm{mf}}(t)=\left(\mathbf{B}+\gamma\mathbf{k}(t)\right)\cdot\mathbf{S}+\gamma\mathbf{s}(t)\cdot\mathbf{K}.\end{equation}
The mean-field dynamics are now given by the Heisenberg equations of motion for the spin operators: $\dot{\mathbf{S}}=i\commute{H_\mathrm{mf}(t)}{\mathbf{S}}$, $\dot{\mathbf{K}}=i\commute{H_\mathrm{mf}(t)}{\mathbf{K}}$, with the replacements $\left<\mathbf{S}\right>_t\approx\mathbf{s}(t)$, $\left<\mathbf{K}\right>_t\approx\mathbf{k}(t)$:
\begin{eqnarray}
\dot{\mathbf{s}}(t) & = &  \left(\mathbf{B}+\gamma\mathbf{k}(t)\right)\times\mathbf{s}(t)\label{eq:MFDE1},\\
\dot{\mathbf{k}}(t) & = &  -\gamma\mathbf{k}(t)\times\mathbf{s}(t)\label{eq:MFDE2}.
\end{eqnarray}
An exact analytical solution to Eqs. (\ref{eq:MFDE1},\ref{eq:MFDE2}) is known.\cite{yuzbashyan:2005a} However, instead of repeating this solution here, we solve Eqs. (\ref{eq:MFDE1},\ref{eq:MFDE2}) by numerical integration for direct comparison with the exact results given above. The mean-field and quantum dynamics are shown in Fig. \ref{fig:MFandQuantumDynamics} for four values of the Zeeman splitting $B=\left|\mathbf{B}\right|$. We compare the two solutions using the correlation function
\begin{equation}
C(t) = \frac{1}{T}\int_{t}^{t+T}dt^\prime\frac{2 \left<S_x\right>_{t^\prime}s_x(t^\prime)}{\left<S_x\right>^2_{t^\prime}+s_x(t^\prime)^2},\label{eq:CDefinition}
\end{equation}
where we average over the time interval $T=0.1 \hbar/\gamma$ to remove rapid oscillations. $C(t)=1$ if the exact solution and mean-field approximation are identical ($s_x(t)=\left<S_x\right>_t$) over the time interval $(t,t+T)$. $C(t)<1$ indicates that the two solutions differ.  While the zero-magnetic-field dynamics appear to be well reproduced by the mean-field approximation, at least at short time scales, the high-field solution decays rapidly, which can not appear in the classical dynamics unless averaging is performed over the initial conditions.\cite{coish:2004a}  There is a partial recurrence of the correlator at a time scale given by the inverse level spacing for the quantum problem $\tau_\mathrm{p} = 2\pi\hbar/\gamma$, but the recurrence is only partial since at this time the quantum and classical solutions have already gone out of phase.  

It is relatively straightforward to understand the difference in the high-field and low-field behavior shown in Fig. \ref{fig:MFandQuantumDynamics}. At zero magnetic field, the total spin $\mathbf{J}\cdot\mathbf{J}\;(\mathbf{J}=\mathbf{K}+\mathbf{S})$ commutes with the Hamiltonian, so if the nuclear spin system begins in an eigenstate of $\mathbf{K}\cdot\mathbf{K}$, only a single frequency exists in the quantum dynamics, corresponding to the difference in energies with $J=K\pm1/2$.\cite{khaetskii:2003a,schliemann:2003a}  Thus, in this case the quantum dynamics corresponds to simple periodic precession, and mimics the classical dynamics for $K>>1$ (see Fig. \ref{fig:MFandQuantumDynamics}(b)).  However, the states of fixed $J$ are manifold degenerate.  If a term is added to the Hamiltonian which does not commute with $\mathbf{J}\cdot\mathbf{J}$ (in this case, the electron Zeeman term $BS^z$), many more frequencies are involved in the quantum dynamics, which can lead to decay in the quantum solution, while the classical solution continues to describe simple electron spin precession (see Fig. \ref{fig:MFandQuantumDynamics}(c)).  In a large magnetic field ($B\gg\gamma|\mathbf{K}_\perp|$), it is straightforward to evaluate the decay in the quantum mechanical solution\cite{coish:2004a},\footnote{The decay formula (Eq. (\ref{eq:decayformula})) is obtained from Eq. (19) of Ref. \onlinecite{coish:2004a} by restoring the formula to dimensionful units and applying the replacements $p\to\cos(\theta_K)$, $A/N\to\gamma$, $N/2\to K$.} 
\begin{eqnarray}
\label{eq:decayformula}
\left<S_+\right>_t &\approx& \left<S_+\right>_0\exp\left\{-\frac{t^2}{2\tau_c^2}+i\left[B+\cos(\theta_K)\gamma K\right]t\right\},\\  
\tau_c &=& \frac{1}{\gamma}\sqrt{\frac{2}{K\left[1-\cos^2(\theta_K)\right]}}. 
\end{eqnarray}
The $x$-component of spin is then given by the real part $\left<S_x\right>_t=\mathrm{Re}\left[\left<S_+\right>_t\right]$. We consider the hyperfine problem with $I=1/2$. When the initial nuclear-spin coherent state is generated by rotating the spins from a fully-polarized state such that $K$ is maximal (as in Ref. \onlinecite{coish:2004a}), we then have $K=N/2$.  In addition, $\gamma = A/N$ and for nuclear spin polarization $p=\cos(\theta_K)\ll 1$ this gives the decay time
\begin{equation}
\label{eq:correlationtime}
\tau_c = 2\frac{\sqrt{N}}{A}.
\end{equation} 
Since the classical dynamics at times $t<\tau_c$ describe simple precession for fixed initial conditions, any decay in the quantum solution signifies a disagreement between the quantum and classical problems.  Thus, the mean-field solution will give an accurate description of the full quantum dynamics only for times $t<\tau_c$, with $\tau_c$ given by Eq. (\ref{eq:correlationtime}).

The crossover from precession to decay of the quantum solution with the addition of a magnetic field suggests that the uniform coupling-constants picture should only be used with caution, since the Hamiltonian in Eq. (\ref{eq:HFHamiltonian}) also does not commute with $\mathbf{J}\cdot\mathbf{J}$ when the coupling constants vary from one nuclear-spin site to the next (as is true in a quantum dot).  Indeed, in the presence of randomly-varying coupling constants, the straightforward mean-field electron-spin dynamics at times $t>\tau_c$ are quantitatively very different from the exact quantum dynamics at weak magnetic fields $B \to 0$.\cite{alhassanieh:2005a}

\section{Conclusions}
We have presented an exact solution for the problem of an electron spin interacting with a large bath of spins with uniform Heisenberg coupling.  This exact solution has been compared to the corresponding mean-field (classical spin) model.  We have seen that the mean-field and quantum solutions show striking agreement at times shorter than the transverse-spin correlation time $\tau_c$, which diverges at zero magnetic field.  This divergence, however, may only be due to the assumption of uniform coupling constants, which is unphysical for a quantum dot with strong confinement.  

In this work we have focused on a comparison of dynamics for fixed initial conditions of the quantum and classical problem.  Some of the quantum behavior, including Gaussian decay, can be recovered with an average over classical solutions.\cite{khaetskii:2002a,merkulov:2002a} An intriguing question therefore remains: How much of the quantum dynamics can be obtained by averaging over classical solutions with different initial conditions?

\emph{Note added}: Recently, a related preprint \cite{zhang:2006a} has appeared in which the authors use the exact solution for uniform coupling constants to evaluate the $z$-component of electron spin, complementing earlier predictions for the bath-polarization dependence of decoherence in single\cite{coish:2004a} and double dots.\cite{coish:2005a} 

\begin{acknowledgments} We acknowledge financial support from the Swiss NSF, the NCCR nanoscience, EU NoE MAGMANet, DARPA, ARO, ONR, JST ICORP, and NSERC of Canada.
\end{acknowledgments}

\pagebreak
\bibliographystyle{apsrev}
\bibliography{icpsproc}
\pagebreak
\listoffigures
\end{document}